\documentclass[a4paper]{article}

\usepackage[english]{babel}
\usepackage[utf8x]{inputenc}
\usepackage[T1]{fontenc}

\usepackage[a4paper,top=3cm,bottom=2cm,left=3cm,right=3cm,marginparwidth=1.75cm]{geometry}

\usepackage{amsmath}
\usepackage{graphicx}
\usepackage{authblk}
\usepackage[colorinlistoftodos]{todonotes}
\usepackage[colorlinks=true, allcolors=blue]{hyperref}

\title{A Review of Beam-Driven Plasma Wakefield Experiments}
\author[1]{A. Aimidula}
\author[2]{P. Zhang}
\affil[1]{School of Physics Science and Technology, Xinjiang University}
\affil[2]{School of Electronic Science and Engineering, University of Electronic Science and Technology of China}

\begin{document}
\maketitle

\begin{abstract}
In the past decades, beam-driven plasma wakefield acceleration (PWFA) experiments have seen remarkable progress by using high-energy particle beams such as electron, positron and proton beams to drive wakes in neutral gas or pre-ionized plasma. This review highlights a few recent experiments in the world to compare experiment parameters and results.
\end{abstract}

\section{Introduction}
In last century, conventional radiofrequency (RF) accelerator has made tremendous contribution to high energy physics and industries such as medical imaging. However, its accelerating gradient is limited to hundreds of MV/m due to electrical breakdown of cavities. In 1970s, Tajima and Dawson \cite{dawson1979} proposed the idea of propagating laser through plasma to create perturbation which laid theoretical foundation of plasma wakefield. Assuming the plasma density is $10^{14} cm^{-3}$, according to Equation \ref{eq:1}, the accelerating gradient is at least ten times higher than conventional accelerator where $E_z$ is longitudinal component of electric field amplitude, $m_e$ is the electron static mass, $c$ is speed of light, $\epsilon_0$ is vacuum permittivity. 
\begin{equation}\label{eq:1}
E_z = \frac{m_{e}c^2}{\epsilon_0}n_{e}^{1/2} \approx 100\sqrt{n_e(cm^{-3})} 
\end{equation}

Depends on plasma disturbance is driven by a charged particle bunch or a laser pulse, plasma accelerator scheme is called plasma wakefield accelerator (PWFA) \cite{joshi2006} or laser wakefield accelerator (LWFA)\cite{Esarey2009}. These two scheme shares many similarities. In this review, we limit the discussion only to PWFA.

The driver beam in PWFA scheme should be ultra-relativistic ($\gamma \ll 1$), normally shorter than plasma wavelength ($\sigma_z < \pi c/\omega_p$) and narrower than plasma column length ($\sigma_r < c/\omega_p$). Here $\gamma$, $\omega_z$, $\omega_r$ and $\omega_p$ are the Lorentz factor, beam RMS bunch length, beam RMS width and plasma frequency respectively. Plasma can be generated by pre-ionizing gas with line-focused laser pulse \cite{Green2014} or directly by the Coulomb field of driver bunch itself \cite{Connell2006}. There is a special scenario worth mentioning that when the bunch density is much higher than the plasma density ($n_b \ll n_p$) where $n_b$ is driver beam density, driver beam will expel plasma electrons on the path completely leaving an ion column behind. This is named blowout regime \cite{Rosenzweig1991,Lu2006}. Inside the blowout regime, there are accelerating and decelerating phase in longitudinal direction. Due to the ion column surrounded by electron sheath, the longitudinal electric field can be tens of GV/m. The scaling relationship is derived as $E_z \propto n_p^{1/2} \propto N/\sigma_z^2$ where $N$ is the number of electrons in bunch. Practically \cite{Chen1985}

\begin{equation}\label{eq:2}
eE_z[MeV/m] \approx 240\times(\frac{N}{4\times10^{10}})(\frac{0.6}{\sigma_z[mm]})^2
\end{equation}

Assuming $N = 2\times10^{10}$, $\sigma_z=30\mu m$, $eE_z$ can be as high as 50 GeV/m. Transversally, \cite{Clayton2002,Muggli2004} have studied matched condition when beam spot size does not have evolution over $z$. The matched condition is $\beta_{matched} = \sqrt{2\gamma}c/\omega_p$ where $\beta$ is beta function of transverse oscillation.

The early remarkable PWFA experiments were carried out at Final Focus Test Beam (FFTB) in SLAC including the famous demonstration of energy doubling test \cite{Blumenfeld2007}. The beam parameters are $\sigma_z < 30\mu m$, $\sigma_r < 10\mu m$. Single bunch was delivered to ionize gas via tunnel ionization to generate plasma. The head of bunch drove the plasma wake, thus losing energy to plasma. Electrons at the tail of the beam witnessed the accelerating wakefield and were accelerated from 42 GeV to 85 GeV in less than one meter of plasma. In addition, a few studies were also established at FFTB including the first demonstration of positron acceleration in plasma \cite{Blue2003}, envelope oscillations of an unmatched electron beam and the concept of beam matching \cite{Clayton2002,Muggli2004} and betatron radiation emitted by off axis electrons in the ion cavity of the wakefield \cite{Wang2002}.

The following three sections will be dedicated to discuss three experiments: FACET in SLAC, AWAKE in CERN and ATF in Brookhaven National Laboratory respectively. Experiment environment and parameters will be discussed as a comparison.

\section{FACET}
FACET is a R\&D facility to study advanced accelerator technologies primarily based on plasma. Topics include high-gradient electron acceleration with narrow energy spread and preserved emittance, efficiency, high-gradient positron acceleration and radiation generation. The first two thirds of LINAC was constructed for FACET while the last third was for x-ray free electron laser the Linac Coherent Light Source (LCLS). 

\begin{figure}
\centering
\includegraphics[width=0.8\textwidth]{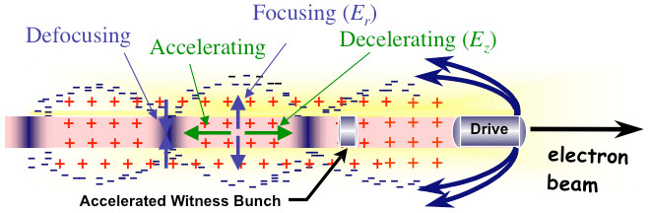}
\caption{\label{fig:pwfa} Schematic picture of beam-driven PWFA with driver beam on the right driving a train of plasma bubble while witness beam following behind at accelerating and focusing phase.}
\end{figure}

FACET driver bunch contains $2\times10^{10}$ electrons which accounts for 3.2 nC. At FACET interaction point (IP) area, driver bunch energy is about 23 GeV with repetition rate of 10Hz. The minimum beam size is designed to be $20\mu m\times 20\mu m\times20\mu m$ which indicates energy densities of 1$MJ/cm^2$ and intensities of $1\times10^{20}W/cm^2$. Parameters are summarized in Table \ref{tab:facetebeam}.

\begin{table}
\centering
\begin{tabular}{l|r}
Parameters & Value \\\hline
beam energy & 23 GeV \\
bunch charge & 3.2 nC \\
beam spot size $\sigma_x$, $\sigma_y$ & 20-30 $\mu m$ \\
bunch length $\sigma_z$ & 20-30 $\mu m$ \\
\end{tabular}
\caption{\label{tab:facetebeam}FACET electron beam parameters}
\end{table}

One PWFA experiment at FACET is double-bunch experiment as shown in Figure \ref{fig:pwfa}. The electron bunch delivered to FACET experimental area has a head-to-tail energy chirp. The W chicane at Sector 10 acts as compressor to compress beam to $\sim 60 \mu m$. The first dipole magnet of the chicane disperses the chirped beam horizontally. Two insertable titanium blades at the high- and low-energy positions of the dispersed bunch were often used as well to split bunch into two. The one with lower energy as driver bunch arrives at experimental area first. M. Litos et al observed energy loss of the driver beam and energy gain of 9 GeV for the witness beam containing 80 pC charge with a 5\% energy spread in 1.2m long plasma\cite{Litos2016}. The efficiency of transferring energy from driver beam to witness beam was also studied. With optimum beam loading, the efficiency can be up to 30\%\cite{Litos2014}.

\begin{figure}
\centering
\includegraphics[width=0.5\textwidth]{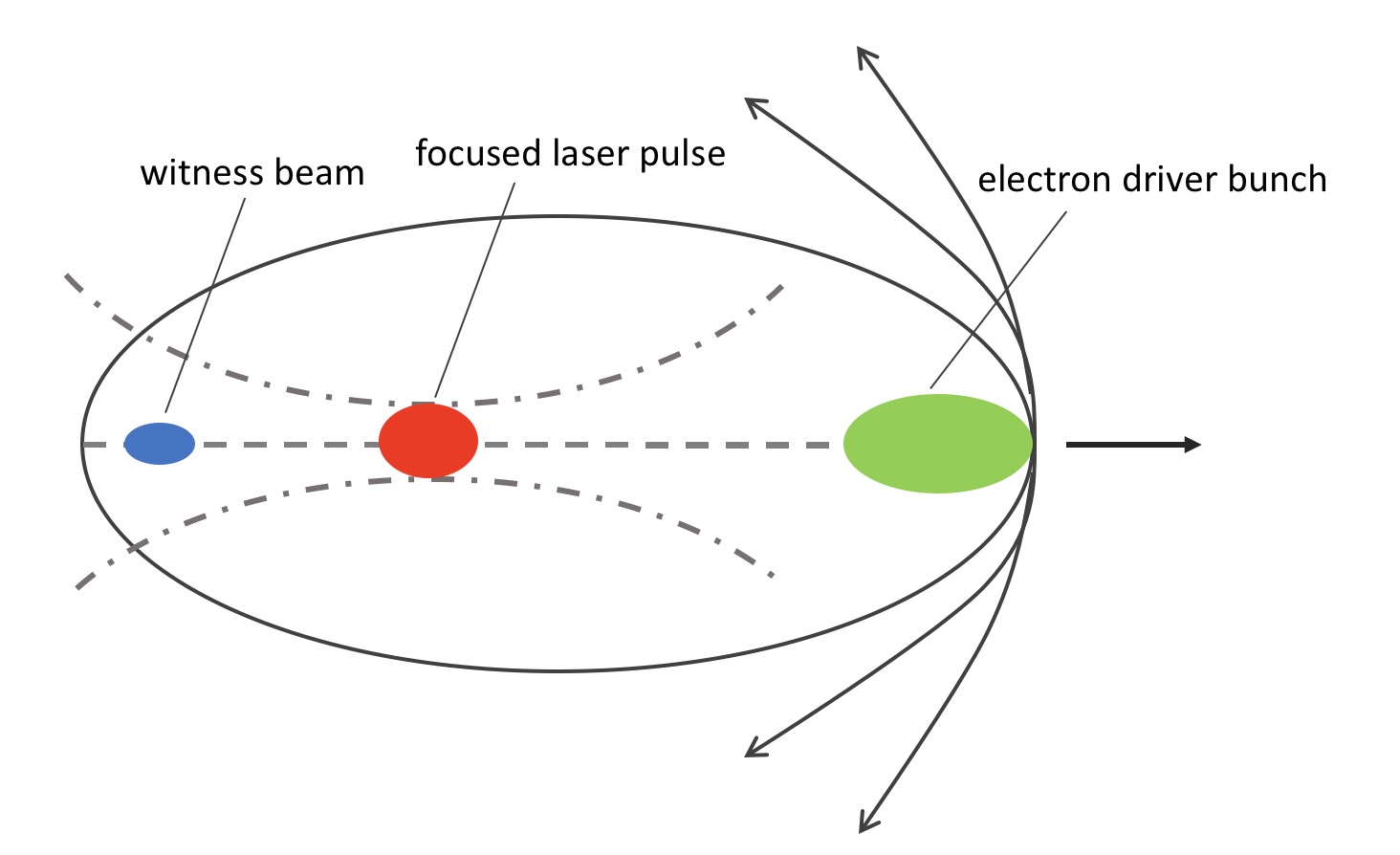}
\caption{\label{fig:trojanhorse} A schematic of "Trojan horse" PWFA. Driver bunch (green dot) drives wake in front. Following behind, synchronized laser pulse (red dot) is focused and ionize He gas to release electrons. These electrons are trapped at the tail of plasma bubble to form the witness beam (blue dot) and accelerated.}
\end{figure}

The PWFA blowout regime has also been tested with positron at FACET. However, positron beam cannot be accelerated in the same blowout regime due to the fact that ion column inside the blowout regime repels positrons. Positron acceleration observed in single positron bunch\cite{Corde2015}. Generation of wakes and acceleration of a distinct positron bunch in a preformed plasma\cite{Doche2017} and a hollow channel plasma\cite{Gessner2016} were also demonstrated.

Besides the double-bunch experiment, there was another PWFA experiment carried out at FACET. The idea was to inject driver beam first to a $\mathrm{H_2}$ and He gas mixture to ionize lower-ionizing-threshold $\mathrm{H_2}$ and create plasma column. Following behind with a certain delay, a ultra-short laser pulse is injected and focused at optimal position to ionize higher-ionizing-threshold gas He. Released electrons will be trapped and accelerated as witness beam. The scheme was nicknamed as "Trojan horse"\cite{Hidding2012,Hidding2013}. According to the simulation and calculation in \cite{Xi2013}, the emittance of witness beam can be as low as $10^{-8} m rad$ which demonstrated to be promising candidate for next generation light source\cite{Hidding2014}. To guarantee the witness beam to be released at appropriate phase, the experiment requires timing synchronization between driver bunch and laser pulse to sub-100-femtosecond level. Thus they proposed "plasma torch" as an intermediate stage which has less restriction on timing\cite{Wittig2015,Wittig2016}. Another important issue in this experiment was the optimization of plasma density or plasma wavelength and the mixing ratio of two gas species. Dark current can spoil witness bunch beam quality and acceleration efficiency in particle beam-driven plasma wakefield accelerators. Hot spots generated by the drive beam and strategies for generating clean and robust, dark current free plasma wake cavities were discussed in \cite{Manahan2016}.

Laser played a key role in this experiment. FACET laser installed in 2013 was a 10 TW Ti:Sapphire system. The majority of laser beam was to pre-ionize gas to generate plasma. A small portion was split out for "Trojan horse" ionization and timing synchronization purpose. Specifically, electro-optic sampling (EOS) was applied for timing experiment\cite{Xi2016}. EOS exploits electro-optic effect that a relativistic electron bunch passes crystals like GaP or ZnTe by a few mm distance. The intense static field modifies refractive index and the induced birefringence rotates the polarization of probe laser beam which can be detected by scanning the relative phase delay. The laser time-of-arrival relative to electron beam was within 100-femtosecond. FACET laser parameters were listed below.

\begin{table}
\centering
\begin{tabular}{l|r}
Parameters & Value \\\hline
wavelength & 800 nm \\
spot size delivered to experimental area & 1 cm \\
pulse duration after compression & 60 fs \\
maximum pulse energy & 500 mJ \\
repetition rate & 1-10 Hz \\
\end{tabular}
\caption{\label{tab:facetebeam}FACET laser beam parameters}
\end{table}

\section{AWAKE}
The Advanced Proton Driven Plasma Wakefield Acceleration Experiment (AWAKE) is a proof-of-principle R\&D experiment at CERN and the first plasma wakefield acceleration experiment driven by proton\cite{Caldwell2013,Assmann2014}. The experiment will be installed in the former CNGS (CERN Neutrinos to Gran Sasso) area. The beam commission was scheduled at the end of 2016. The first demonstration of proton-driven PWFA was scheduled at the end of 2017.

\begin{figure}
\centering
\includegraphics[width=0.8\textwidth]{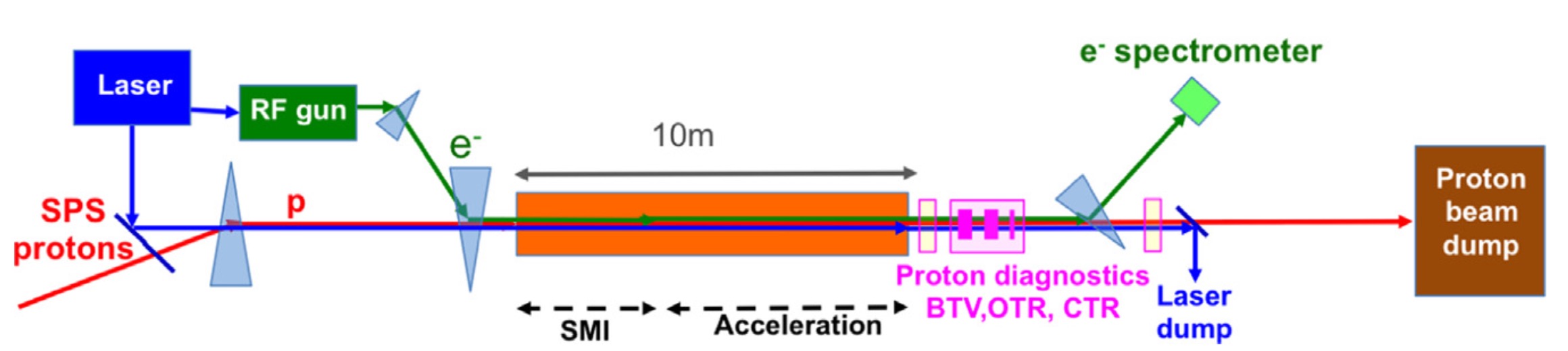}
\caption{\label{fig:awakebaseline}Schematic design of the AWAKE baseline experiment as shown in \cite{Gschwendtner2016}}.
\end{figure}

The baseline design of AWAKE experiment is shown in Figure \ref{fig:awakebaseline}. Proton beam from SPS and laser are injected from left. The plasma length is 10m. The downstream diagnostics include beam position monitors(BTV), OTR and CTR.

The mission of AWAKE is to explore the self-modulation of long proton bunches (~1mm) in plasma as a function of beam and plasma parameters including radial modulation and seedings of the instability. The second project is to measure the accelerated beam parameters with external injection. The study of injection dynamics and the production of multi-GeV electron bunches is also planned. The last goal is to develop long, scalable and uniform plasma cell to accelerate proton bunches as advanced accelerator technology.

The AWAKE experiment requires short bunches with very high intensity from CERN SPS. Beam operating in this condition is unstable and difficult to control parameters from shot to shot. To reduce bunch length as much as possible, bunches should be rotated in the longitudinal phase space by a quarter of the synchrotron period. This is achieved by reducing and then sharply increasing the RF voltage. The shortest achievable bunch length is determined by the maximum RF voltage and the smallest longitudinal emittance. With this method, proton beam at AWAKE can have $3\times10^{11}$ protons per bunch, normalized transverse emittance of 1.7 mm mrad, bunch length of 9 cm and peak current of 60 A. Parameters are list in Table \ref{tab:awake}.

\begin{table}
\centering
\begin{tabular}{l|r}
Parameters & Value \\\hline
proton beam energy & 400 GeV \\
protons per bunch & $3\times10^{11}$ \\
bunch length $\sigma_z$ & 0.4 ns \\
bunch size $\sigma_{x,y}$ & 200 $\mu m$ \\
repetition rate & 0.03 Hz \\
plasma type & vaporized rubidium \\
plasma density & $7\times10^{14}\mathrm{cm^{-3}}$ \\
\end{tabular}
\caption{\label{tab:awake}AWAKE proton beam and plasma parameters}
\end{table}

One important topic at AWAKE is self-modulation instability. It has been discussed since decades ago\cite{Max1974}. Most of plasma wakefield experiments operate on short bunches and large plasma densities to obtain large amplitude wakefields due to scaling of wave breaking field. AWAKE experiment will use long proton bunches and thus be in self-modulated plasma wakefield regime. According to \cite{Kumar2010}, the condition that plasma return current cannot intersect driver bunch put the upper limit on plasma density. The condition is satisfied when the bunch transverse size $\sigma_r$ is smaller than the cold plasma collisionless electron skin depth $c/\omega_{pe}$ where $c$ is speed of light and $\omega_{pe}$ is plasma angular frequency. The dissatisfaction would cause current filamentation instability which prevents the efficient excitation of plasma wakefields\cite{Allen2012}.

\section{ATF}
The Accelerator Test Facility (ATF) at Brookhaven National Laboratory has been a user facility of accelerator research for decades. ATF provides high-brightness 80 MeV electron beam and 1 TW picosecond $\mathrm{CO_2}$ laser. These are powerful tools to study electron-beam-driven PWFA which is a remarkable member among ATF projects. Relevant projects include but not limit to developing capillary discharge as plasma source\cite{Pogorelsky2003}, demonstrating phasing between the longitudinal and transverse components of the wakefields acting on driver electron bunch\cite{Yakimenko2003}, demonstrating a high-gradient, controlled acceleration of a short electron bunch trailing the driver electron bunch in high-density plasma\cite{Kallos2008}, observing the current filamentation instability of an electron beam propagating in plasma with tunable densities\cite{Allen2012}, demonstrating seeding of the self-modulation instability resulting in the periodic modulation of the electron bunch and studying the resonance multi-bunch PWFA as well as technique to generate multiple bunches\cite{Muggli2008}.

\begin{figure}
\centering
\includegraphics[width=0.8\textwidth]{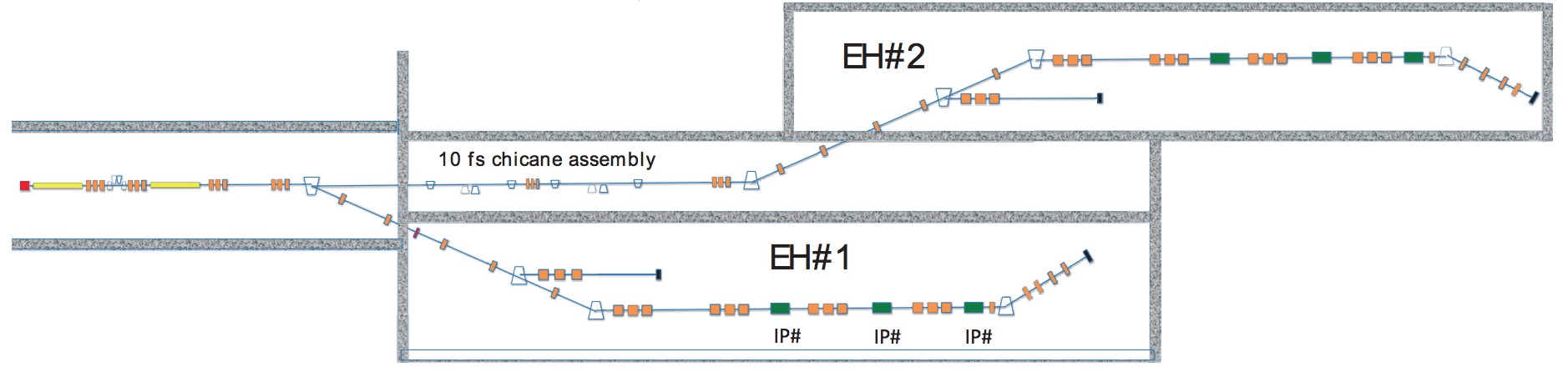}
\caption{\label{fig:atf}Design of ATF II beamline from \cite{Fedurin2015}}.
\end{figure}

ATF accelerator is composed of a BNL/SLAC/UCLA style S-Band photoinjector. Following that are two traveling wave linac capable of boosting electron energy to 50-80 MeV. Then electron beam is transported through an initial drift section which has two electromagnetic quadrupole triplets and an electromagnetic chicane which the former ones allow beam tuning to deliver a tailored beam to one of two experimental halls and the latter one is to compress beam. At the end, a dipole redirects beam to experimental area. Beam is transported to experimental halls via "dog-leg" dispersion sections.

The energy, charge and emittance of current ATF beam barely meet the minimum requirements to drive nonlinear blowout regime PWFA as in Equation \ref{eq:3} where $N_b$ is total electron number, $n_b$ is bunch density, $\sigma_r$ is the beam RMS size, $n_0$ is the plasma density and $\lambda_p$ is plasma wavelength. 

\begin{equation}\label{eq:3}
n_b/n_0 > 1, \sigma_r \ll \lambda_p, N_b/n_0\lambda^3_p \gg 1
\end{equation}

Quasi nonlinear plasma wakefield acceleration has been studies based on these conditions \cite{Rosenzweig2012,Barber2014}. Usually PWFA operates in nonlinear or blowout regime. However, linear regime also has some advantages such as the possibility for resonant excitation of wakefields through the use of pulse trains. Experiments have been carried out at ATF to use electron pulse trains to observe resonant wakefield excitation and to observe strong focusing of a matched pulse train through the plasma.

\begin{table}
\centering
\begin{tabular}{l|r}
Parameters & Value \\\hline
beam energy & 300 MeV \\
maximum charge & $3 nC$ \\
bunch current & 100 A \\
normalized emittance & 1 mm-mrad \\
compressed bunch length & 3 $\mu m$ \\
\end{tabular}
\caption{\label{tab:awake}AWAKE proton beam and plasma parameters}
\end{table}

ATF has planned to upgrade to Phase II when electron beam energy is improved to 300 MeV and $\mathrm{CO_2}$ laser's peak power increases to 100 TW. This upgrade will allow more opportunities of PWFA research. While upgrading $\mathrm{CO_2}$ laser will enable laser wakefield accelerator, improving electron beam energy enables exploration of nonlinear regime of PWFA. The required increase of peak current can be achieved by focusing and compressing beam more. The upgrade with better compressor and a new photocathode electron gun will ensure the satisfaction of blowout conditions.

\section{Summary}
In conclusion, three beam-driven plasma wakefield acceleration experiments at FACET, AWAKE and ATF are reviewed. While AWAKE is under ongoing development, FACET and ATF are planning and commissioning Phase II upgrade. Their experiment environment and beam parameters are compared. We also list the experimental characteristics and goals. This could be insightful guidelines for future beam-driven PWFA planning.

\bibliographystyle{alpha}
\bibliography{sample}

\end{document}